\documentclass[12pt,preprint]{aastex}

\newcommand{\Mpch}{h^{-1}\mathrm{Mpc}}

\def\eg{{\it e.g.}}
\def\ltsima{$\; \buildrel < \over \sim \;$}
\def\simlt{\lower.5ex\hbox{\ltsima}}   
\def\gtsima{$\; \buildrel > \over \sim \;$}
\def\simgt{\lower.5ex\hbox{\gtsima}}
\def\hide#1{}

\begin{document}

\title{Is the Concentration of Dark Matter Halos at Virialization
  Universal ?}  

\author{Massimo Ricotti\altaffilmark{1},
Andrew Pontzen\altaffilmark{2}, and
Matteo Viel\altaffilmark{2,3}} \altaffiltext{1}{Dept. of
Astronomy, U. of Maryland, College Park, MD 20742}
\altaffiltext{2}{Institute of Astronomy, Madingley Road, Cambridge CB3 0HA} 
\altaffiltext{3}{INAF - Osservatorio Astronomico di Trieste, Via
  G.B. Tiepolo 11, I-34131 Trieste, Italy} 

\begin{abstract}
Several recent studies suggest a correlation between dark matter halo
mass and the shape of the density profile.  We re-analyze simulations
from Ricotti (2003) in which such a correlation was proposed. We use a
standard analysis of the halo density profiles and compare the old
simulations to new ones performed with Gadget2, including higher
resolution runs.  We confirm Ricotti's result that, at virialization,
the central log slopes $\alpha$, at 5\%-10\% of the virial radius are
correlated with the halo mass and that the halo concentration, $c$, is
a universal constant. Our results do not contradict the majority of
published papers: when using a split power law to fit the density
profiles, due to the $\alpha-c$ degeneracy, the fits are consistent
with halos having a universal shape with $\alpha=1$ or $1.5$ and
concentrations that depend on the mass, in agreement with results
published elsewhere.

Recently, several groups have found no evidence for convergence of the
inner halo profile to a constant power law.  The choice of a split
power law parameterization used in this letter is motivated by the
need to compare our results to previous ones and is formally valid
because we are not able to resolve regions where the slope of the
fitting function reaches its asymptotic constant value. Using a
non-parameterized technique, we also show that the density profiles of
dwarf galaxies at $z \sim 10$ have a log slope shallower than $0.5$
within $5$\% of the virial radius.
\end{abstract}
\keywords{Galaxies: formation; Methods: N-body simulations; Cosmology: theory.}

\section{Introduction}\label{sec:introduction}

It is widely believed that dark matter halos have a universal density
profile.  In the last decade the debate focused largely on whether the
slope of the density profile, $\alpha$, approaches values close to $1$
or $1.5$ at small radii. The concentration, $c$, of dark halos
supposedly incorporates all the physics: the dependence on the
cosmological model, the halo mass and redshift of virialization.
However, the determination of the inner slope of halos is uncertain
due to the scatter of halo shapes and a well known degeneracy between
$\alpha$ and $c$ when fitting the density profiles.  Indeed, only a
few published simulations have sufficiently high resolution to
partially resolve this degeneracy and these simulations only cover a
limited range of halo masses (typically Milky-Way type or cluster type
halos) at redshift $z=0$.

In this paper, we re-analyze data from N-body simulations performed in
\cite{2003MNRAS.344.1237R}, henceforth R03, and compare them to a new
set of simulations performed using Gadget2 \citep[][]{Springel:2005mi}
at the same and higher resolution than in R03.  The R03 results are of
particular interest because, in R03, it was first found that the log
slope of the inner parts (at 10\% of the virialization radius for just
virialized halos) of the dark matter density profile, $\alpha$, varies
with the mass of the halo. This result may ease some tension between
theory and observations. Indeed, for low mass halos ($<10^{9}
M_{\sun}$), which should correspond to dSph galaxies
\citep[\eg,][]{RicottiG:05, ReadPV:06} and perhaps some Low Surface
Brightness (LSB) galaxies \citep[\eg,][]{2002A&A...385..816D}, R03
finds $\alpha <1$ at 10\% of the virial radius (where $r_{vir} \sim
1$~kpc).  In R03, Milky-Way type halos with mass $M \sim
10^{12}~M_{\sun}$ are well fitted by a NFW profile with $\alpha \sim
1$ \citep{1996ApJ...462..563N} and there is also some evidence that
higher mass systems ($ > 10^{15} M_{\sun} $) have steeper cusps,
$\alpha > 1$.  This is consistent with results in
\cite{1999MNRAS.310.1147M}, although this latter paper also proposes
$\alpha\sim 1.5$ for rather small, galaxy mass halos ($ \sim 10^{11}
M_{\sun}$).
It is important to clarify that the quoted values of $\alpha$ in the
inner part of the halos depend on the resolution of the simulations
(\eg, in R03 is fixed to be about 10\% of the virial radius).  There
is no reason to believe that $\alpha$ converges to any asymptotic
value as suggested by the arguments in R03 and and by high-resolution
simulation of Milky-Way size halos
\citep{2004MNRAS.349.1039N,Graham:2005xx}. Note that the result in R03
does not depend on this assumption. In R03 we show that NFW profile
does not provide good fits for small mass halos at $z\sim 10$ by
comparing the shapes of the circular velocities of just virialized
halos with widely different masses (hence, virialized at different
redshifts). Our conclusion is that density profiles do not have an
universal shape as often assumed before. A theoretical interpretation
of this result is also proposed based on previous work by
\cite{2000ApJ...538..528S} in which a simple relationship that relates
$\alpha$ to the slope of the power spectrum of initial density
perturbations is provided. Our results are consistent with this
simple scaling relationship, suggesting that the halo shape at a given
mass or spatial scale depends on the the slope of the power spectrum
at that scale. This result is based on formal fitting of the circular
velocities of the halos with generalized NFW
profiles. \cite{Colin:2003jd} performed similar simulations and failed
to find shallow cores in a similar mass range.  Nevertheless, similar
correlations have been found by others,
\citep[e.g.][]{2000ApJ...529L..69J, 2001ApJ...563..483T, Chen:2004bm}.
More recent work by \citet{Graham:2005xx} and
\citet{2005ApJ...624L..85M} also find a correlation between halo mass
and the shape of the density profile. In this case they parameterize
this correlation in terms of the S\'{e}rsic index, $n$, rather than a
central log-slope dependence. In \citep{2004MNRAS.349.1039N,
2005ApJ...624L..85M} is argued that this parameterization provide a
better fit of high-resolution halos than a split power-law.

The main motivation of this letter is to understand whether the R03
results are in contradiction with previous works and clarify if the
discrepancy can be attributed to the method of analysis, the N-body
code used by R03, or insufficient resolution of the
simulation. Ricotti used a $P^3M$ N-body integrator and analyzed the
data using circular velocities instead of density profiles. Here we
adopt the widely used ``tree'' N-body integrator Gadget2, with $256^3$
and $512^3$ particles and we develop a quantitative method to analyze
the density profiles using a standard $\chi^2$ minimization technique.

This letter is organized as follows. In Section \ref{sec:data}, we
describe the set of simulations from which data have been
used and the procedures we have
adopted and developed for analyzing individual halos; in Section
\ref{sec:results} we present the results; finally we conclude in
Section~\ref{sec:concl--disc}.

\section{Simulation Data and Analysis}\label{sec:data}

All simulations used and referred to in this work have identical
cosmological parameters: $\Omega_{m,0}=0.3$, $\Omega_{\Lambda,0}=0.7$,
$n_s=1$, $\sigma_8=0.91$ and $h=0.7$.  R03 uses a $P^3M$ integrator
\cite[][]{1996ApJ...470..115G} while our new simulations employ the
tree code Gadget2 \cite[][]{Springel:2005mi,2001NewA....6...79S}. The
general notation we use to label each simulation is: Run-L, where Run
= R03, GR03, G256 and G512 describe a different simulation and L =
1,32,256 refers to the box size (i.e., $1,32,256\Mpch$). The runs
``R03'' refer to the original R03 simulations, ``GR03'' use the
initial conditions in R03 but are re-run using Gadget2, ``G256'' and
``G512'' are new runs using Gadget2 with $256^3$ and $512^3$
particles, respectively. The purpose of running the new simulations
GR03, G256 and G512 was to ensure the results in R03 were not affected
by any irregularity in the simulation method employed. GR03 checks for
problems with the R03 simulation parameters and code; G256 checks the
initial conditions and G512 checks for resolution related issues.  

The redshift of analysis in all $1\Mpch$ simulations is $z=10$. The
larger box sizes are analyzed when the clustering of the most massive
halos is similar, which turns out to be $z=3$ for $L=32\Mpch$ and
$z=0$ for $L=256\Mpch$.
We extracted the halos using a Friends-of-Friends algorithm with
linking length $l=0.2$ (chosen by analogy with the spherical collapse
model). We follow the iterative method of \cite{2002MNRAS.332..325P}
to ensure that the extracted halos are bound. At each stage the total
energy of each particle is calculated. Particles which do not appear
to be bound are excluded from the potential calculation in the next
stage. 
An accurate determination of the halo centers is important, since from
a miss-centered halo we would deduce a systematically flattened
profile.  We use three centering methods: \textit{i) Density
Maximum}. An algorithm for finding the density maximum uses an
adaptive grid of cubes; where the number of particles inside a cube
exceeds a certain threshold, the cube is subdivided and the process
iterates. \textit{ii) Shrinking Sphere Center of Mass}, as in
\citet{2003MNRAS.338...14P}. Here we calculate the center of mass
(COM) and then iterate, including at each stage only particles within
some sphere around the previously calculated center. \textit{iii)
Potential Minimum}. The particle with the minimum potential is chosen
as the center. Generally we find that the potential minimum lies
within the convergence radius of the shrinking sphere center. All
three alternative centering procedures are used; if one procedure does
not produce results within the convergence radius of each of the
others, the halo is flagged as unusable for spherically averaged
processing.  We use the convergence criterion of
\cite{2003MNRAS.338...14P} to determine a central region where the
density distribution is unreliable. This yields a convergence radius
inside which the profile is not considered during the fitting
procedure.

We fit the density profiles using a standard $\chi^2$ minimization
routine with a generalized NFW profile \citep{1997ApJ...490..493N} in
which the concentration $c$ and inner slope $\alpha$ are free
parameters: $\rho \propto x^{-\alpha}(x+1)^{-3+\alpha}$, where
$x=r/r_s$ and $r_s=r_{vir}/c$.  It is obvious that this
parameterization of the profile should be regarded valid only between
the virial radius and the convergence radius which, for all the halos
considered in this work, is a constant fraction $\sim 5-10$\% of the
virial radius.  In this paper, we use the Poisson error for our
$\chi^2$ minimization procedure.  It is widely noted
\cite[e.g.][]{2000ApJ...535...30J,2004MNRAS.349.1039N} that errors
introduced by deviations from an ``idealized'' equilibrium profile
(i.e. substructure, asphericity and irregularities) are likely to be
at least as important as Poisson errors.

\section{Results}\label{sec:results}

\begin{figure}
  \includegraphics[width=0.3\textwidth]{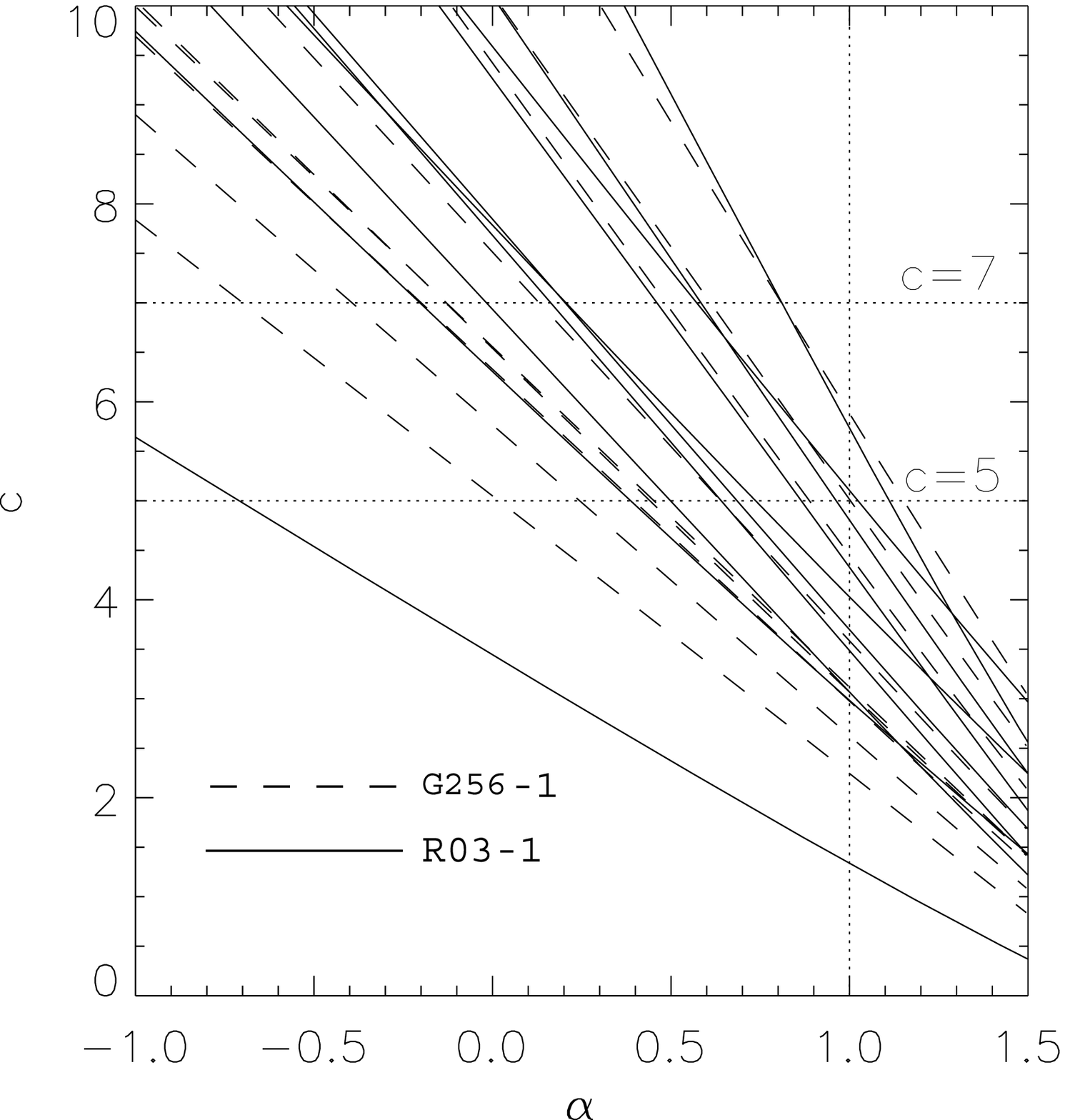}
  \includegraphics[width=0.3\textwidth]{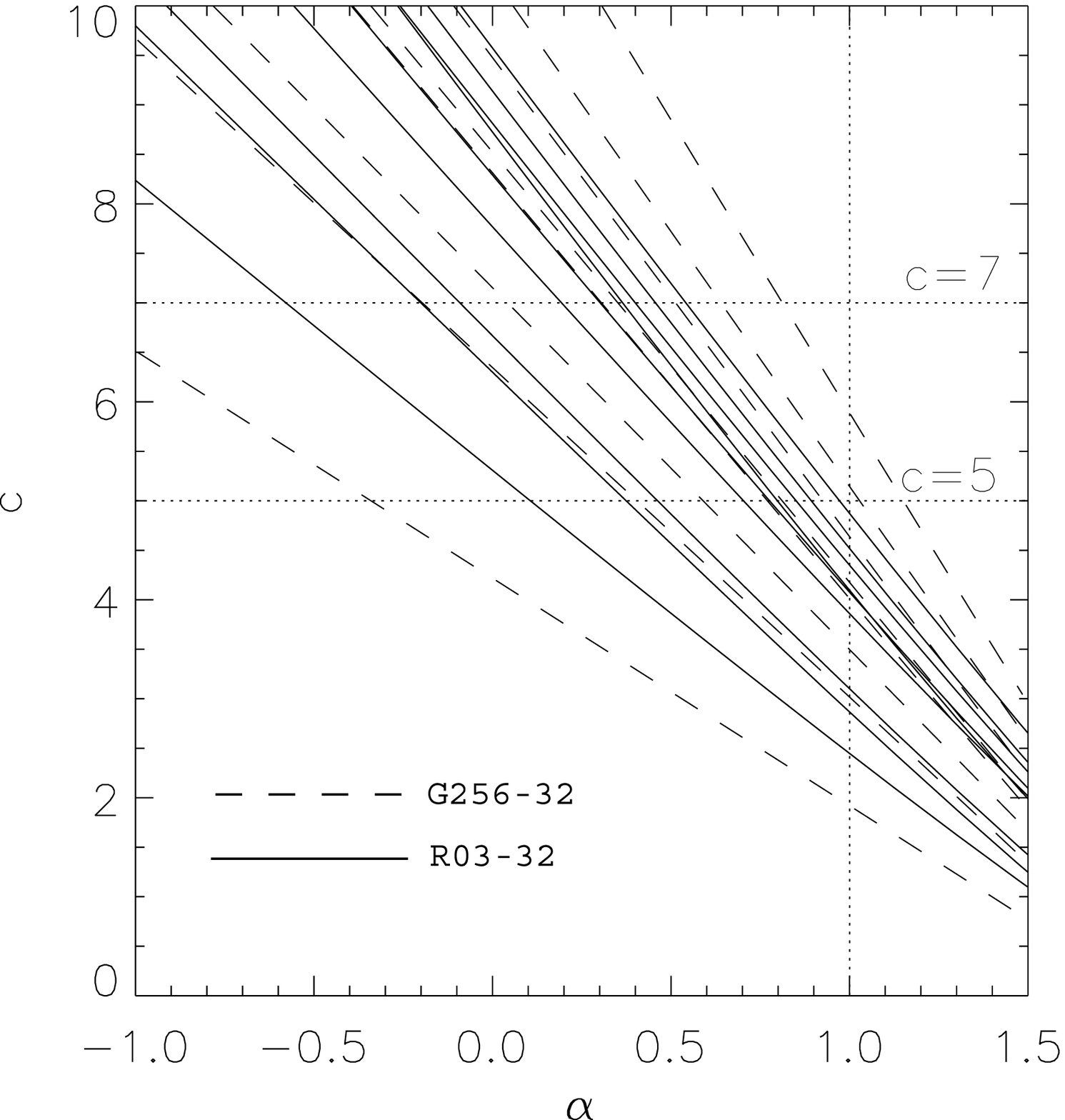}
  \includegraphics[width=0.3\textwidth]{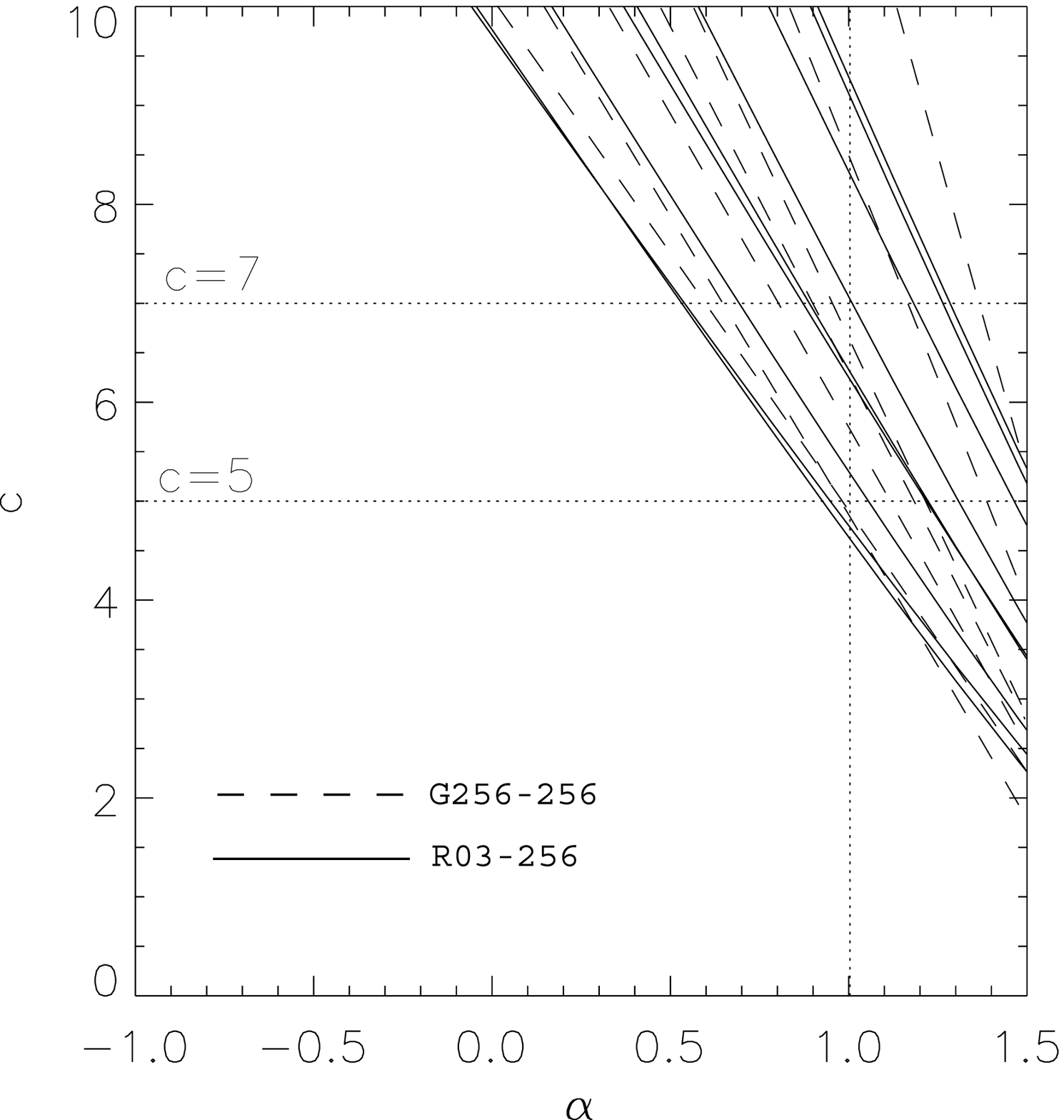}
\caption{From left to right: line of minimum $\chi^2$ for fits (by
  density) to the first 10 R03 halos for box sizes $L=1,32,256\Mpch$.
  The results from the Gadget2 runs (GR03, G256 and G512) are in
  excellent agreement; G256 results are overplotted as an example of
  these.  The intersection of each line with the vertical or
  horizontal dotted lines gives the fit constrained to an NFW profile
  or a fixed $c$ split power-law profile respectively. While $\chi^2$
  varies along each line, it does so by a factor of only a few;
  moreover the mean $\chi^2$ obtained for constraining a number of
  halo fits along one axis or the other is almost identical (see
  Table \ref{tab:fitresults-big}).
}\label{fig:results-bestfitline}
\end{figure}

When fitting the split power law to a halo, there is a degeneracy
between $c$ and $\alpha$: the model is effectively constrained not to
a point but to a line through the $c$-$\alpha$ plane, along which
$\chi^2$ varies by only a small factor $\sim 2$ from its minimum to
the edges of the region of interest $0<\alpha<1.5$. The location of
locus of minima varies widely from halo to halo and may select
non-physical configurations with $\alpha<0$. Perpendicular to this
line, however, the value of $\chi^2$ increases rapidly.  In Figure
\ref{fig:results-bestfitline} left panel, we illustrate this
degeneracy for the 10 most massive halos in R03-1. By constraining
either $c$ or $\alpha$ arbitrarily, each line can be reduced to a
point, given by the intersection with a horizontal or vertical line
respectively.  When constraining $\alpha = 1$, which gives NFW fits,
the best-fit $c$ lies in the range $2 \simlt c \simlt 5$.  For $c=5$,
the best fit $\alpha$ lies in the range\footnote{ Note that in
\cite{2004MNRAS.353..867R} it is found that a concentration parameter
of $c=5$ is a better estimate than the original value $c=7$ in R03. We
therefore adopt this value for obtaining numerical results (Table
\ref{tab:fitresults-big}).} $0 \simlt \alpha \simlt 1$. The results
from the Gadget2 runs (GR03, G256 and G512) were in excellent
agreement with the R03 runs.

By using an analysis of the peak circular velocity, R03 found that
recently virialized halos are well described by a constant
concentration $c=5$. This value was then used to break the profile
fitting degeneracy, giving values of $\alpha$ which agree with the
theoretical prediction of \cite{2000ApJ...538..528S} for a range of
halo masses from the various box sizes -- for more details see the
original R03 paper.  When one uses circular velocity
fitting, as in R03, with a Poisson assumption we obtain a smaller
$\Delta\alpha=0.1$ range.  If we use density fitting with
the NFW $\alpha=1$ constraint, illustrated by a vertical line in
Figure \ref{fig:results-bestfitline}, the first 40 halos
in R03-1 give $c=3.7\pm 1.4$. This closely matches the results given
in \cite{Colin:2003jd} -- who analyze their $1\Mpch$ box at redshift
$z=3$ instead of $z=10$ -- once corrected for redshift as $c \propto
(1+z)^{-1}$ as in \cite{2001MNRAS.321..559B}. The remainder of the
$1\Mpch$ results are shown in Table \ref{tab:fitresults-big}; all seem
to be consistent and allow for fitting using either constraint.  The
increased mean $\chi^2$ in the G512-1 simulation occurs because the
Poisson errors are reduced (more particles per bin) while the
systematics due to departures from the fitted split power-law profile
are constant, since these arise through ``physical'' processes which
are not greatly changed in magnitude by the change in resolution.
\label{sec:asympt}
\nocite{2004MNRAS.353..867R}

We now consider whether there is any evidence for differences between
the $1\Mpch$ and boxes and their $32$ and $256\Mpch$ counterparts.
The line of minimum $\chi^2$ for the 10 most massive regular halos in
R03-32 and R03-256 is plotted in the middle and right panels of Figure
\ref{fig:results-bestfitline} respectively. Taking the 40 most massive
halos and their intersections with the constraint $c=5$ gives
$\alpha=0.6\pm0.4$ ($L=32\Mpch$) and $\alpha=1.4\pm0.4$
($L=256\Mpch$). The NFW $\alpha=1$ constraint gives $c=3.6\pm1.4$ and
$c=9.1\pm2.6$ respectively. Comparing with the $L=1\Mpch$ results,
there is a marginal difference in the $L=32\Mpch$ results, and a clear
difference in the $L=256\Mpch$ results. The comparison simulations
G256-32, G256-256 and GR03-256 show the same trend with box size (see
Table~\ref{tab:fitresults-big}).

It is important to note that neither the NFW profile plotted in Figure
3 of R03, nor the profiles with a shallower central slope $\alpha \sim
0.4$ have reached their asymptotic values at the convergence radii.
The innermost resolved radii, however, satisfy $\beta=d\ln
v_{cir}/d\ln r = 1 - \alpha/2 > 0.5$, which cannot be obtained by a
true NFW profile (which must have $\beta \le 0.5$). Unfortunately,
using the density profile fitting we find that direct estimation of an
individual profile's logarithmic slope yields results too noisy to
perform direct analysis near the convergence radius, even with the
higher resolution G512 simulations.

\subsection{A Non-parametric Method}

\begin{figure}
\plotone{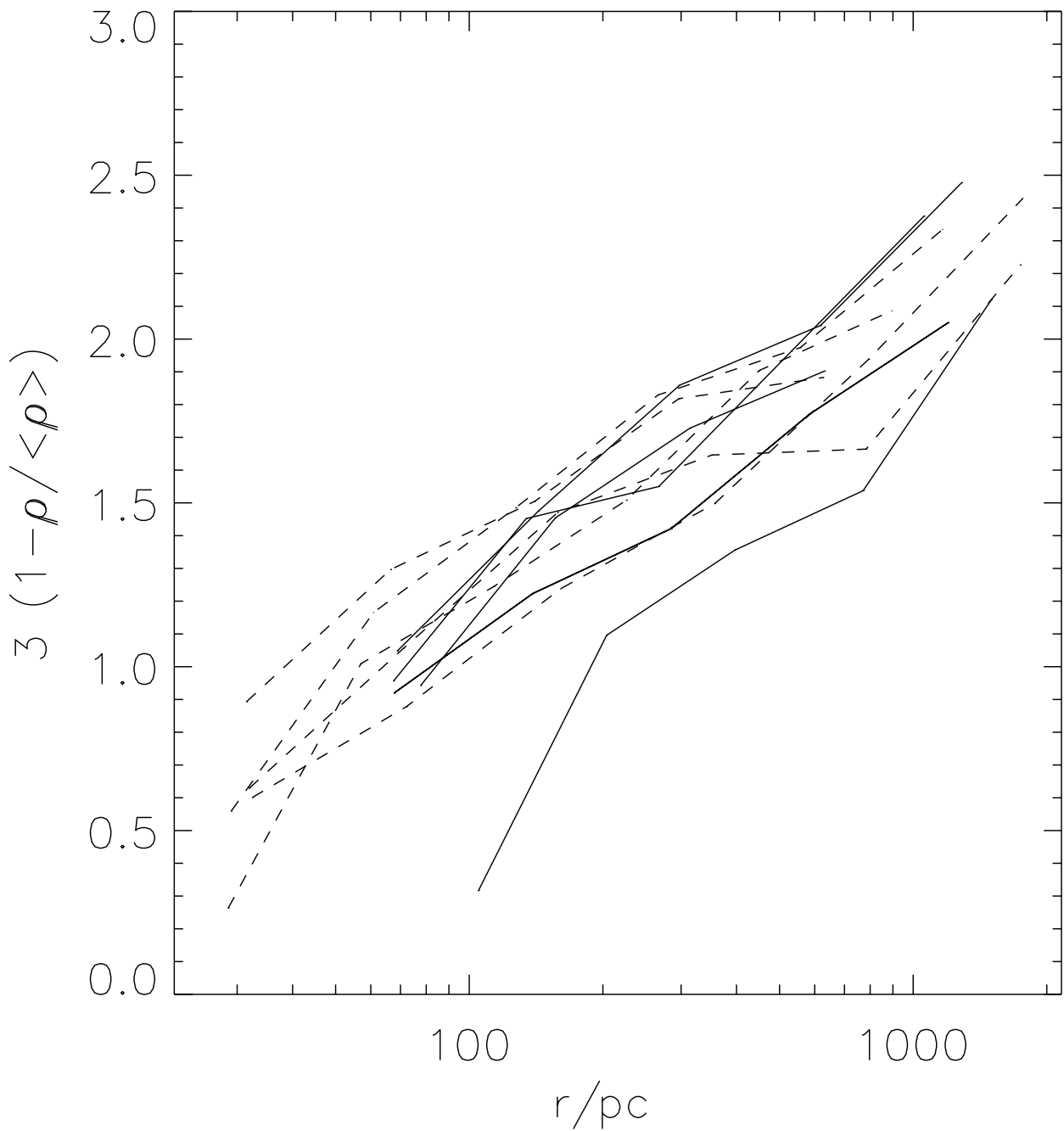}
\caption{The first five most massive G512 (dashed) and R03 halo
    $\gamma=3(1-\rho/\bar{\rho})$ profiles. Each profile is plotted
    only exterior to its convergence radius (which are smaller for the
    G512 simulations). $\gamma$ would converge to $\alpha$ in the case
    of the split power law profile. There does not appear to be any
    evidence for convergence to a particular value.}\label{fig:gamma}
\end{figure}

To circumvent the problems of the fitting degeneracy, we follow the
method of \citet{2004MNRAS.349.1039N}. This procedure is similar to the
one used by R03 where the circular velocities where analyzed instead
of the density profiles.
We make the assumption that the total mass interior to the convergence
radius, $r_c$, is representative of the ``physical'' halo, even if the
inner profile is wrong.  This being the case, $ \gamma =
3\left(1-\rho(r)/\bar{\rho}(r)\right)$ gives a robust limit to the
asymptotic power law of the halos: we cannot have a steeper than $\rho
\propto r^{-\gamma}$ cusp.  The $\gamma$ -- radius relation is plotted
for some sample halos in Figure \ref{fig:gamma}. The G512-1 halos are
resolved to smaller radii than the R03 run. The sets are in
approximate agreement where their data overlaps, but the innermost
point of the G512 profiles have $\gamma<1$, suggesting that the
profile interior to $r_c$ would not be well described by NFW but
probably in agreement wiht the latest high-resolution simulation
results \citep{2004MNRAS.349.1039N, Graham:2005xx}. However, $r_c$ is
probably smaller than the cores which may have been observed in LSB
(see e.g. \cite{2003ApJ...588L..21K}, \cite{2001ApJ...552L..23D} and
\cite{Goerdt:2006rw}). That said, \cite{2004MNRAS.353..867R} have
shown that the density profiles do provide good fits to the kinematic
data from the Draco Local Group dSph.


%
\section{Conclusions \& Discussion}\label{sec:concl--disc}

In R03, Ricotti have claimed that central slopes of dark matter
density profiles (at a fixed fraction of the virial radius, dictated
by the resolution of the simulation) are correlated with halo mass and
that the halo concentration at the redshift of virialization is a
universal constant. This result is consistent with the finding of recent
high-resolution simulations at $z=0$ \citep{2004MNRAS.349.1039N,
Graham:2005xx} in which is found that the slope of the density profile
in halos become flatter with decreasing radius without converging
to any asymphotic value.

It's often a misconception that the results in R03 contradict several
published papers. For example, Merritt et al. (2005, 2006) found that
the average inner slope for four dwarfs at $z=0$ with mass $10^10$
M$_\odot$ is $-1.24$ as compared with $-1.17$ for 8 clusters. This
does not contradict our result because clusters and dwarfs virialize
at very different redshifts. The virial radius and concentration of
dwarfs at $z=0$ has increased by a factor $z_f+1 \sim 10$ from the
redshift of their formation, $z_f$. Hence, the inner slope we measure,
{\it i.e}., at 5\%-10\% of the virial radius in $10^8$ M$_\odot$ halos
at $z=10$, should be compared to the slope at 0.5\%-0.1\% of the
virial radius at $z=0$ in comparable mass dwarfs. In addition, in the
abstract of the same paper, they confirm a dependence of the halo
profile on the halo mass, which is the main result in R03.  Diemand,
Moore, Steidel (2005), found that in minihalos of mass
$10^{-6}$ M$_\odot$ at $z=26$ the density profile has inner slope
$-1.2$ at 10\% of the virial radius. But the number of particles in
the three halos they look at is a factor of $10-100$ smaller than in
our simulations. In addition the power spectrum of density
fluctuations they use differs from ours as they use a power law with
an exponential cut off at 0.6 parsec. It is not clear how the cut off
in the power spectrum can affect the density profile. For example,
contrary to our simulations, their halos do not have substructure.

Here, we have re-analyzed the results presented in R03 and run new
simulations showing that R03 results are reproducible by using
different N-body codes, higher-resolution simulations, and different
analysis techniques. Part of the tension between R03 and previous
results may be attributed to the method of analysis of the
profiles. When fitting the density profiles, as is done in most of
previous works, the fitting degeneracy between $\alpha$ and $c$ does
not allow one to understand whether is the slope $\alpha$ or the
concentration $c$ that depends on the mass of the dark halo. If one
favors the \cite{2000ApJ...538..528S} scaling arguments which express
a relation between $\alpha$ and the initial power spectrum (thus,
between $\alpha(r)$ and the enclosed mass $M(<r)$), it is found that
the concentration of halos at the redshift of virialization is a
universal constant.  However, Figure~\ref{fig:results-bestfitline}
shows that the R03 results can be re-parameterized by taking a
different cut through the degeneracy, leading to the more widely
accepted notion of the NFW concentration depending on the halo mass
and epoch \cite[e.g][]{2001MNRAS.321..559B,Colin:2003jd}.  R03, by
analyzing the circular velocities of small mass galaxies at $z \sim
10$, finds some evidence for cusps shallower than $\alpha=1$ at radii
$<10$\% of the virial radius (in contradiction with NFW profiles).
Here, using a different non-parameterized analysis we also find some
evidence for a flatter than $\alpha=1$ power-law of the density
profile within $5-10$\% of the virial radius in galaxies with mass
$\sim 10^9$ M$_\odot$ at $z \sim 10$. Note that the scaling argumennt
in \cite{2000ApJ...538..528S} advocated by R03 suggests that the
density profile does not converges toward any given asymptotic value
of the logarithmic slope but rather becomes gradually flatter toward
the center. This is in good agreement with the results of recent
high-resolution simulations at $z=0$ \citep{2004MNRAS.349.1039N,
Graham:2005xx}. However, the present work does not have sufficient
resolution to investigate this hypothesis and compare the goodness of
different shapes for the fitting function.

\section*{Acknowledgments}
We thank Justin Read for feedback. The simulations were run on the
COSMOS supercomputer at the DAMTP, Cambridge. COSMOS is a UK-CCC
facility which is supported by HEFCE and PPARC.

\clearpage

\clearpage

\begin{deluxetable}{l|lll|lll}
\tabletypesize{\scriptsize}
\tablecaption{\label{tab:fitresults-big}}
\def\arraystretch{1.0}
\tablecolumns{7} 
\tablehead{ 
\colhead{Sim} & \multicolumn{3}{c}{Constraint: $c=5$} & 
\multicolumn{3}{c}{Constraint: $\alpha=1$} \\ 
\cline{2-4} \cline{5-7} \\ 
\colhead{} & \colhead{Mean}   & \colhead{Variance} & 
\colhead{$\left<\chi^2\right>$}    & \colhead{Mean}   & \colhead{Variance} & \colhead{$\left<\chi^2\right>$}}
\startdata
\cutinhead{$L=1 Mpc$ at $z=10$}
 R03-1 & $\alpha=0.3$ & 0.6 & 11 & $c=3.7$ & 1.4 & 12\\
 GR03-1 & $\alpha=0.4$ & 0.7 & 15 & $c=3.8$ & 1.5 & 13\\
 G256-1 & $\alpha=0.4$ & 0.5 & 11 & $c=3.4$ & 1.0 & 12\\
 G512-1 & $\alpha=0.2$ & 0.6 & 120 & $c=3.5$ & 1.2 & 111\\
\cutinhead{$L=32 Mpc$ at $z=3$}
 R03-32 & $\alpha=0.6$ & 0.3 & 11 & $c=3.9$ & 1.4 & 12\\
 G256-32 & $\alpha=0.6$ & 0.4 & 15 & $c=3.8$ & 1.6 & 8\\
\cutinhead{$L=256 Mpc$ at $z=0$}
  R03-256 & $\alpha=1.4$ & 0.4 & 11 & $c=9.1$ & 2.6 & 12\\
  GR03-256 & $\alpha=1.3$ & 0.4 & 15 & $c=8.8$ & 3.0 & 14\\
  G256-256 & $\alpha=1.6$ & 0.4 & 16 & $c=9.2$ & 3.6 & 15\\
\enddata\tablecomments{Results for fitting a split power law profile to the 40
  most massive halos in each simulation, with constraints as follows:
  \textit{(1)} Fix $c=5$ (as in Ricotti \& Wilkinson (2004)),
  $\beta=3$ and measure $\alpha$; \textit{(2)} Fix $\alpha=1$ (as
  NFW/C04), $\beta=3$ and measure $c$. There is no detectable
  correlation between the halo masses and any of the parameters
  analyzed here. Therefore the results are a true reflection of the
  ensemble of halos and their intrinsic scatter. Note that the
  $\langle\chi^2\rangle$ for both constraint are comparable: there is
  no evidence that one is to be preferred over the other. The higher
  $\langle \chi^2 \rangle$ in the G512 halos is to be expected due to
  the constant systematics but smaller Poisson errors. The values
  obtained for the NFW concentration are in agreement with those
  obtained in other published results, once corrected for redshift
  (see e.g. Bullock et al. 2001)
  }
\end{deluxetable}

\clearpage

\clearpage

\end{document}